\documentclass[prb,twocolumn,showpacs,preprintnumbers,amsmath,amssymb]{revtex4}
\usepackage{graphicx}
\usepackage{dcolumn}
\usepackage{bm}

\begin{document}

\title{Critical behavior of density of states near Fermi
energy in low-dimensional disordered metals}

\author{E. \c{S}a\c{s}{\i}o\u{g}lu$^{1,2}$}\email{e.sasioglu@fz-juelich.de}
\author{S. \c{C}al{\i}\c{s}kan$^{2}$}\email{scaliskan@fatih.edu.tr}
\author{M. Kumru$^{2}$}\email{mkumru@fatih.edu.tr}

\affiliation{$^{1}$Institut f\"{u}r Festk\"{o}rperforschung,
Forschungszentrum J\"{u}lich, D-52425 J\"{u}lich, Germany\\
$^{2}$Department of Physics, Fatih University, TR-34500,
B\"{u}y\"{u}k\c{c}ekmece, \.{I}stanbul, Turkey}

\date{\today}

\begin{abstract}
We study the effect of electron-electron interaction on the
one-particle density of states (\emph{DOS})
$\rho^{(d)}(\epsilon,T)$ of low-dimensional disordered metals near
Fermi energy within the framework of the finite temperature
conventional impurity diagram technique. We consider only
diffusive limit and by a geometric re-summation of the most
singular first order self-energy corrections via the Dyson
equation we obtain a non-divergent solution for the \emph{DOS} at
low energies, while for higher energies the well-known
Altshuler-Aronov corrections are recovered. At the Fermi level
$\rho^{(d)}(\epsilon,T=0)\rightarrow 0$, this indicates that
interacting disordered two- and quasi-one-dimensional systems are
in insulating state at zero temperature.  The obtained results are
in good agreement with recent tunneling experiments on
two-dimensional GaAs/AlGaAs heterostructures and
quasi-one-dimensional doped multiwall carbon nanotubes.
\end{abstract}


\maketitle

\section{Introduction}
In the last three decades, a great deal of progress has been made
towards revealing the behavior of electrons in a random potential.
Efforts have led to a detailed understanding of the
low-temperature properties of the weakly  disordered systems,
i.e., systems for which $k_F  l \gg 1$ where $k_F$ is the Fermi
wave number and $l$ is the elastic mean free path.
\cite{Abrikosov} This understanding has been embodied in weak
localization theory and disorder enhanced electron-electron
(\emph{ee}) interaction effects.\cite{Bergmann,Altshuler} The
interplay of \emph{ee} interaction and random impurity potential
on the transport and thermodynamic properties of disordered
systems has been studied intensively. In particular, treatment of
the problem within perturbation theory lead to the understanding
of the anomalous logarithmic decrease in conductivity with
decreasing temperature in two-dimensional ($2d$) electron gas,
negative magnetoresistance observed in many $2d$ and $3d$ systems
as well as the depression in the \emph{DOS} near the Fermi
energy.\cite{Altshuler,Fukuyama,Ramakrishnan}

Extensive experimental and theoretical studies have shown that
interaction effects are enhanced by disorder and generally result
in a decrease of the \emph{DOS} near the Fermi level. Studies on
the theoretical side have, for the most part, concerned two
extreme limits. In the limit of strong disorder, this decrease
takes the form of a complete gap in the \emph{DOS} at the Fermi
energy.\cite{gap_1,gap_2,gap_3,gap_4} It is known that this
\emph{Coulomb gap} can turn a highly disordered pure metal into a
poorly conducting insulator. In the opposite limit, i.e., the
diffusive limit, in a  pioneering paper Altshuler, Aronov and Lee
(AAL) treated the $2d$ disordered electron problem within the
perturbation theory to lowest order in interaction
strength.\cite{AAL} The authors showed that interaction effects in
a 2$d$ disordered metal lead to the development of a logarithmic
singularity in the one-particle \emph{DOS}, $\delta
\rho^{(2)}(\epsilon)\sim\ln(|\epsilon|\tau) $ near the Fermi
energy $\epsilon_F$, where $\tau$ and $\epsilon$ are the impurity
scattering time and the energy of the electron measured from the
Fermi level, respectively. Such effects become even more stronger
in quasi-$1d$ disordered metals, $\delta\rho^{(1)}(\epsilon)\sim
-(|\epsilon|\tau)^{-1/2}$.\cite{Altshuler} Unlike the
low-dimensional systems, the quantum corrections to the \emph{DOS}
in $3d$ is rather small, $\delta \rho^{(3)}(\epsilon)\sim
\sqrt{|\epsilon|\tau}$ giving rise to cusp at the Fermi
energy.\cite{Altshuler_Aronov} Extension of the AAL theory to the
ballistic limit shows that the interaction effects give rise to
non-trivial corrections to the corresponding physical properties
of the disordered systems also in this regime.
\cite{bl_1,bl_2,bl_3,bl_4,bl_5,bl_6,bl_7,bl_8} The reduction in
the one-particle \emph{DOS} near the Fermi energy results in
suppression of the tunneling conductance at small voltages, a
phenomenon commonly known as the zero-bias anomaly (ZBA). Several
groups have performed tunneling measurements on $3d$ and
quasi-$2d$ systems since early 1980's and observed the predicted
dependence of the tunneling conductance with
voltage.\cite{tunnel_1,tunnel_2,tunnel_3,tunnel_4,tunnel_5,tunnel_6}

Recently the problem of \emph{ee} interaction in disordered metals
received considerable  interest due to the  discovery of the
unexpected metallic state in high-mobility two-dimensional
semiconductors  by Kravchenko \textit{et
al.},\cite{2d_metal_1,2d_metal_2} and development of the new
experimental techniques such as time domain capacitance
spectroscopy (TDCS).\cite{tdcs_1} The existence of a metallic
state in 2$d$ with finite conductivity at zero temperature is in
conflict with the conventional weak localization theory, which
predicts that even negligible amount of disorder in low
dimensional systems ($d\leq 2$) localizes electrons at
sufficiently low temperatures. Thus, in spite of several
theoretical proposals, the metallic state in 2$d$ is one of the
puzzling phenomena that is still waiting for an adequate
description.\cite{2d_metal_3,2d_metal_4,2d_metal_5,2d_metal_6,2d_metal_7,2d_metal_8,2d_metal_9,2d_metal_10}
On the other hand, TDCS is appeared to be a very useful technique
for detection of the tunneling current in regimes difficult to
access by conventional methods, and, thus allows the quantitative
comparison of the existing theories with
experiments.\cite{tdcs_2,tdcs_3,tdcs_4} Using TDCS Chan \textit{et
al}., for the first time measured the entire voltage dependence of
the tunneling conductance of a two-dimensional electron system in
a GaAs/AlGaAs heterostructure for different electron
densities.\cite{tdcs_2,tdcs_3} The authors observed the expected
logarithmic Coulomb anomaly only in the case of a small
suppression of the tunneling current. However, for large
suppressions corresponding to small electron densities the
functional form of the ZBA vs bias voltage was significantly
deviating from the predictions of the AAL theory, especially  in
the regime of very small voltages.

The critical behavior of the \emph{DOS} for $\epsilon \rightarrow
0$ in reduced dimensions which is not accessible within the
first-order perturbation theory, is of great interest to
understand the low-temperature transport and thermodynamic
properties of the disordered metals. In this respect an initial
attempt was made by Finkelstein using field theoretic
renormalization group theory, who found that
$\rho^{(2)}(\epsilon)\sim \epsilon^{1/4}$ as  $\epsilon
\rightarrow 0$ for $2d$ systems.\cite{Finkelstein} Since then
there has been a lot of attempt to study energy and temperature
dependence of the \emph{DOS} around the Fermi level employing
different
methods.\cite{Oppermann,Fukuyama2,Castellani,Belitz,Dobrosavljevic,Levitov,Kopietz,Kamenev,Rollbuhler,Oreg}
Kopietz has considered a $2d$ system and by re-summing the most
singular contributions to the average \emph{DOS} via a
gauge-transformation obtained that $\rho^{(2)}(\epsilon) \sim
C|\epsilon|/e^4$ for $\epsilon \rightarrow 0$, where where $C$ is
a dimensionless constant and $e$ is the charge of the
electron.\cite{Kopietz} Kamenev and Andreev  using Keldysh
$\sigma-$model  derived a non-perturbative result for the
\emph{DOS} of quasi-$2d$ systems.\cite{Kamenev} Rollb\"{u}hler and
Grabert extended this work to quasi-$1d$ systems including
additionally the inter–electrode interactions and obtained a
non-divergent solution for the \emph{DOS} at low energies, that
recovers the $(|\epsilon|\tau)^{-1/2}$ behavior for higher
energies.\cite{Rollbuhler} It should be emphasized that in reduced
dimensions, in contrast to first-order perturbation theory (AAL
theory), all these different methods yield a non-divergent
solution for the \emph{DOS} around Fermi level with a power-law
behavior, whereas for higher energies results of AAL theory is
recovered.

The aim of the present work is a detailed study of the critical
behavior of \emph{DOS} around Fermi level in low-dimensional
disordered metals  within the diagrammatic perturbation theory.
This technique, in contrast to above mentioned non-perturbative
schemes, provides a mathematically clear and transparent framework
in studying impurity problems in condensed-matter physics. In the
present work we consider only diffusive regime and go beyond the
first-order perturbation theory. We show that a geometric
re-summation of the most singular first-order self-energy
corrections via the Dyson equation gives a non-divergent solution
for the \emph{DOS} at low energies, while for higher energies the
obtained expressions are reduced to the predictions of the AAL
theory. At zero temperature in both dimensions the \emph{DOS}
vanishes at the Fermi energy. In spite of good agreement between
present approach and above mentioned non-perturbative treatments
for the \emph{DOS} at small corrections (higher energies) an
essential difference appears in the asymptotic energy dependence
of the \emph{DOS}. The remaining of the paper is organized as
follows. In Sec.~\ref{section2} high-order perturbation
corrections to the \emph{DOS} is calculated and compared with AAL
theory. In Sec.~\ref{section3} we dwell on the zero-bias anomaly
of the tunneling conductivity and make a qualitative comparison of
the obtained results with  recent tunneling experiments.
Section~\ref{section4} gives the conclusions.

\begin{figure}[t]
\begin{center}
\includegraphics[scale=0.6]{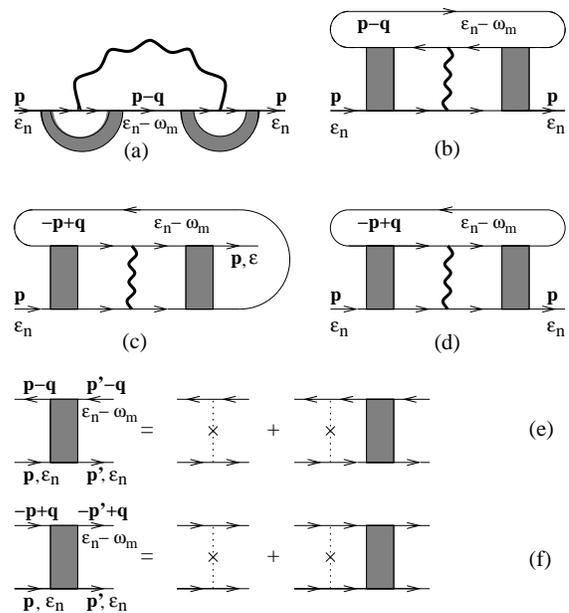}
\end{center}
\vspace*{-0.4cm} \caption{ Diagrams  for the  calculation of
$\Sigma({\bf p},i\epsilon_n)$; (a), (c) exchange diagrams for the
diffusion and Cooper  channels, respectively; (b), (d) Hartree
diagrams for the diffusion and Cooper  channels, respectively. The
thick wavy lines denote the dynamically screened Coulomb
interaction. (e), (f) ladder series for the diffusion and Cooper
channels. Here the dashed line with cross denotes the impurity
scattering.} \label{fig1}
\end{figure}
\begin{figure}[t]
\begin{center}
\includegraphics[scale=0.54]{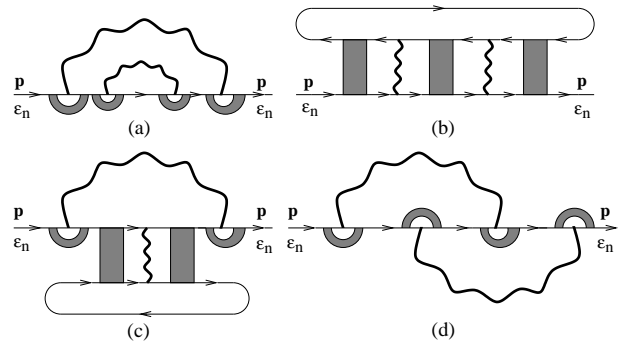}
\end{center}
\vspace*{-0.4cm} \caption{Some particular diagrams for
second-order in the Coulomb interaction corrections to the
self-energy.} \label{fig2}
\end{figure}

\section{High-order perturbation corrections to the \emph{DOS}}
\label{section2}

As it is well known the main contributions to the physical
properties of disordered systems in the weak localization theory
are connected with two singularities: First appears in the
diffusion propagator, characterizing an electron-hole pair with
small difference of the momenta ${\bf q}$ and of the energies
$\omega$ (Diffusion pole). Other singularity is due to propagation
of electron-electron pair with small sum of the momenta ${\bf q}$
and small difference of the energies $\omega$ (Cooper pole). In
weak disorder case, corrections to the \emph{DOS} can be obtained
through the self-energy $\Sigma({\bf p},i\epsilon_n)$. First order
in Coulomb interaction contributions to the self-energy are
illustrated in Fig.~\ref{fig1}. Higher order in Coulomb
interaction also gives contributions to the self-energy. Some
particular diagrams concerning the second order self-energy
contributions are presented in Fig.~\ref{fig2}. However, it can
easily be shown that the ratio of the second order contributions
to the first order ones is found to be
$\lambda_\rho\sqrt{\vert\epsilon\vert\tau}\ll1$, where $
\lambda_\rho $ is the dimensionless constant of
interaction.\cite{Altshuler_Aronov} Therefore we can restrict
ourself to the first order contributions to the self-energy and
neglect small contributions from the higher order self-energies.

To begin with, consider the exchange interaction between the
electrons in diffusion channel depicted in Fig.~\ref{fig1}(a).
This process gives rise to a significant contribution to the
\emph{DOS} and the expression for the self-energy can be written
\begin{eqnarray}
\Sigma^{D}({\bf p},i\epsilon_n)&=&T\sum_{\omega_{m}}
\int\frac{d^{d}q}{(2\pi)^{d}}\:V({\bf q},i\omega_m)\nonumber \\
&&\times \: G_{0}^{A}({\bf p} - {\bf q} ,i\epsilon_n- i\omega_m) \nonumber \\
&&\times \: \gamma^{2}( {\bf
q},i\omega_m)\theta(\epsilon_n(\omega_m-\epsilon_n))
\label{self-energy}
\end{eqnarray}
where $G^{A}_{0}$ is the \emph{bare} temperature Green's function
(GF) for electrons averaged over the impurity potential,
$\epsilon_n=\pi T(2n+1)$ and $\omega_m=2\pi T m $ are the
Matsubara frequencies at temperature $T$. $\gamma({\bf
q},i\omega_m)$ is the sum of the impurity ladders in
Fig.\:\ref{fig1}, which has a diffusion pole under condition
$\vert {\bf q} \vert l\ll1$ and $\vert\omega \vert\tau\ll1$. The
expression for $\gamma({\bf q},i\omega_m)$  is given by
\begin{equation}
\label {ladder_series}
\gamma({\bf q},i\omega_m)=\theta(\epsilon_n(\epsilon_n-\omega_m))+
\frac{\theta(-\epsilon_n(\epsilon_n-\omega_m))}{\tau(\vert\omega
\vert +Dq^2)}
\end{equation}
with $D=\frac{v_F^2 \tau}{d}$ being the diffusion coefficient for
a $d$ dimensional system. $V({\bf q},i\omega_m)$ in
Eq.~(\ref{self-energy}) is the dynamically screened Coulomb
potential. Within the random-phase approximation the $V({\bf
q},i\omega_m)$ takes the following form
\begin{eqnarray}
\label {screened_potential}
V({\bf q},i\omega_m) &=&\frac{2\pi e^2}{\vert
q\vert+\kappa_2\frac{Dq^2}{\vert\omega\vert+Dq^2}},\:\:\:\:\:\:\:\:\:\:\:\:\:\:\:\:\:\:\:\:\:
(d=2)\nonumber \\
&=&\frac{e^2}{e^2\rho_0^{(1)}\frac{Dq^2}{\vert\omega\vert+Dq^2}+\ln^{-1}\frac{1}{q^2a^2}},
\: (d=1)
\end{eqnarray}
where  $\kappa_2=2\pi e^2\rho_0^{(2)}$ being the inverse screening
length for a $2d$ system and $a$ is the transverse size of the
quasi-$1d$ system. $\rho_0^{(d)}$ being the \emph{DOS} of a
non-interacting electron gas which is given by
\begin{equation}
\label {pure_dos}
\rho_0^{(1)}=\frac{1}{2\pi v_F},
\:\:\:\:\rho_0^{(2)}=\frac{m}{2\pi\hbar^2}, \:\:\:\:
\rho_0^{(3)}=\frac{m p_F}{2\pi^2\hbar^2}
\end{equation}
Since the \textbf{q}-integral in Eq.~(\ref{self-energy}) is
dominated by the diffusive pole of the impurity ladders then
within the accuracy of our calculation, for small ${\bf q}$ and
$\omega$, Eq.\:(\ref{self-energy}) can be rewritten in the
following suggestive form
\begin{equation} \label{self-energy2}
\Sigma^{D}({\bf p},i\epsilon_n)\cong \alpha(\epsilon,
T)G_0^{A}({\bf p},i\epsilon_n)
\end{equation}
with
\begin{equation}
\alpha(\epsilon,T)=T \sum_{\omega_{m}}
\int\frac{d^{d}q}{(2\pi)^{d}}\: \gamma^{2}( {\bf q},i\omega_m)
V({\bf q},i\omega_m) \label{alpha}
\end{equation}

The \emph{DOS} $\rho^{(d)}(\epsilon, T)$ of a $d$-dimensional
system is defined in terms of the \emph{total} retarded GF,
$G^{R}({\bf p},i\epsilon_n)$
\begin{equation}
\rho^{(d)}(\epsilon, T)=-\frac{1}{\pi}\textrm{Im}\int\frac{d^d
p}{(2\pi)^d}G^{R}({\bf
p},i\epsilon_n)_{i\epsilon_n\rightarrow\epsilon}
\label{dos1}
\end{equation}
It is well known that the temperature Green's function coincides
with the retarded one at discrete points on the positive imaginary
semiaxis, \textit{i.e.}, $G(\epsilon_n)=G^{R}(i\epsilon_n)$ at
$\epsilon_n>0$. According to the Dyson equation the total Green's
function $G^{R}({\bf p},i\epsilon_n)$ including electron
correlations in diffusion channel is given by
\begin{eqnarray}
G^{R}({\bf p},i\epsilon_n)&=&\frac{1}{\big[G_{0}^{R}({\bf
p},i\epsilon_n)\big]^{-1}
-\Sigma^{D}({\bf p},i\epsilon_n)}\nonumber\\
&=&\sum_{n=0}^{\infty}\big[G_{0}^{R}({\bf
p},i\epsilon_n)\big]^{n+1} \big[\Sigma^{D}({\bf
p},i\epsilon_n)\big]^{n}
\label{Dyson}
\end{eqnarray}
Substituting Eq.~(\ref{self-energy2}) into Eq.~(\ref{Dyson}) and
utilizing Eq.~(\ref{dos1}) the \emph{DOS} takes the following
form\cite{Nakhmedov}
\begin{equation}
\rho^{(d)}(\epsilon, T)=\rho_0^{(d)}-\frac{1}{\pi}\textrm{Im}
\sum_{n=1}^{\infty}A_n\big[\alpha(\epsilon, T)\big]^n
\label{dos2}
\end{equation}
where
\begin{equation}
A_n=\int\frac{d^d p}{(2\pi)^d}\big[G_0^R({\bf
p},i\epsilon_n)\big]^{n+1}\big[G_0^A({\bf p},i\epsilon_n)\big]^{n}
\label{An}
\end{equation}
It is easy to see that $n=0$ term in Eq.~(\ref{dos2}) is equal to
$\rho_0^{(d)}$, thus the bare \emph{DOS} is distinguished. Upon
performing this integration we find
\begin{equation}
A_n=-\rho_0^{(d)}2\pi i\,\tau^{2n}\frac{n(2n-1)!}{(n!)^2}
\label{An2}
\end{equation}
Substituting Eq.~(\ref{An2}) into Eq.~(\ref{dos2}) and taking the
sum over $n$, one obtains the total contribution to the \emph{DOS}
from the diffusion channel
\begin{equation}
\rho^{(d)}(\epsilon, T)=\rho_0^{(d)}-\rho_0^{(d)}\textrm{Im}\bigg
\{\frac{\beta}{{\sqrt{1-i\beta}}\big(1+{\sqrt{1-i\beta}}\big)}\bigg\}
\label{dos3}
\end{equation}
where $\beta=4\tau^2\alpha(\epsilon, T)$. Note that we use the
expression
$\ln\big[1+\sqrt{1+x^2}\:\big]=\ln2-\sum_{n=1}^{\infty}(-1)^n\frac{(2n-1)!}{(n!)^22^{2n}}x^{2n}$
in evaluating the sum in Eq.~(\ref{dos2}).

For a short range (static) Coulomb interaction the equation above
can be simplified considerably. In this case Coulomb potential
depends neither \textbf{q} nor $\omega$ and  thus the integration
in Eq.~(\ref{self-energy2}) is straightforward, in which
calculation for two and one dimensions gives
\begin{eqnarray}
\alpha(\epsilon,T)
&=&-\frac{\pi\lambda_\rho}{8\epsilon_F\tau^3}+i\frac{\lambda_\rho}
{4\epsilon_F\tau^3}\ln\bigg[\frac{1}{2\tau(|\epsilon|,T)}\bigg],
(d=2)\nonumber \\
&=&\frac{\lambda_\rho}{4\tau^2\sqrt{2\tau(|\epsilon|,T)}}(1+i),
\:\:\:\:\:\:\: \:  \: \: \: \: \: \: \: (d=1)
\label{alpha2}
\end{eqnarray}
where $\lambda_\rho=\rho_0^{(d)}V(0,0)$ and $\epsilon$ is the
energy reckoned from the Fermi level.

If we write $\beta\equiv \beta_{R}+i\beta_{I}$ where $\beta_{R}$
and $\beta_{I}$ are real and imaginary parts, respectively, and
then the Eq.~(\ref{dos3}) takes the following form
\begin{equation}
\rho^{(d)}(\epsilon,
T)=\frac{\rho_0^{(d)}}{\sqrt{2}}\frac{\sqrt{1+\beta_R+\sqrt{(1+\beta_R)^2+\beta_I^2}}}{\sqrt{(1+\beta_R)^2+\beta_I^2}}
\label{dos4}
\end{equation}
Using Eqs.~(\ref{alpha2}) and (\ref{dos4}) one obtains the
following expressions for the \emph{DOS} of $2d$ and quasi-$1d$
systems.
\begin{equation}
\rho^{(2)}(\epsilon, T)\simeq\frac{\rho_0^{(2)}}{\sqrt{1+
\frac{2\lambda_\rho}{\epsilon_F\tau}\ln\bigg[\frac{1}{2\tau(|\epsilon|,T)}\bigg]}}
\label{dos_2_d}
\end{equation}
\begin{equation}
\rho^{(1)}(\epsilon,
T)\simeq\frac{\rho_0^{(1)}}{\sqrt{1+\frac{2\lambda_\rho}
{\sqrt{2\tau(|\epsilon|,T)}}}} \label{dos_1_d}
\end{equation}
It follows from Eqs.~(\ref{dos_2_d}) and (\ref{dos_1_d}) that for
small corrections one recovers the results of the AAL theory.
\begin{equation}
\rho^{(2)}(\epsilon, T)\simeq \rho_0^{(2)}
\bigg(1-\frac{\lambda_\rho}{\epsilon_F\tau}\ln\bigg[\frac{1}{2\tau(|\epsilon|,T)}\bigg]\bigg)
\label{dos_2_d_aal}
\end{equation}
\begin{equation}
\rho^{(1)}(\epsilon, T)\simeq \rho_0^{(1)}
\bigg(1-\frac{\lambda_\rho} {\sqrt{2\tau(|\epsilon|,T)}}\bigg)
\label{dos_1_d_aal}
\end{equation}

Consideration of the direct process in diffusion channel depicted
in Fig.~\ref{fig1}(b) yields similar expression for the
\emph{DOS}. Thus, the $\lambda_{\rho}$ in Eq.~(\ref{alpha2})
should be replaced by $\lambda_{\rho}^{D}=\rho_0^{(d)}\big[V(
0,0)-2\overline{V({\bf p}'- {\bf p}'',0)}\big]$ where the bar over
Coulomb potential corresponding to the Hartree diagram denotes
averaging over the Fermi surface and the factor $2$ appearing
because electrons with both spin orientations contribute to the
Hartree correction. Note that Hartree term involves zero energy
and large momentum transfers.

In above expressions the constant $\lambda_\rho$ is the only
unknown parameter that can not be derived in a general way. For a
dynamically  screened Coulomb interaction within some
approximations the $\lambda_\rho$ can be cast into the following
form
\begin{eqnarray}
\lambda_\rho&=& \frac{1}{2}\ln{\bigg[\frac{(|\epsilon|,T)}{\hbar\tau\big(D\kappa_2^2\big)^2}\bigg]}-\frac{3}{2}F,
\:\:\:\:\:\:\: (d=2)\nonumber\\
&=&\frac{a\kappa_3}{\sqrt{\pi}}\ln^{1/2}\bigg[\frac{D\kappa_3^2}{(|\epsilon|,T)}\bigg]-\frac{3}{2}F,
\:\:\: (d=1)
\label{effective_constant}
\end{eqnarray}
where $\kappa_3=\sqrt{4\pi e^2\rho_0^{(3)}}$ and the first terms
represent the exchange contribution to the effective interaction
constant $\lambda_\rho$  in diffusion channel while the second
term ($\frac{3}{2}F$) is associated with the Hartree contribution
in the same channel.  The specific nature for Coulomb interaction
in low-dimensional systems manifest itself only in a logarithmic
dependence of the constant $\lambda_\rho$ on $\epsilon$ and $T$.
In contrast to exchange process the Hartree or direct contribution
to the $\lambda_\rho$ is relatively small ($F\ll1$), in both
dimensions the evident expression of the parameter $F$ is
logarithmic\cite{rosenbaum}
\begin{eqnarray}
F&=&\frac{2}{x^2}\ln(1+x^2),
\:\:\:\:\:\:\:\:\:\:\:\:\:\:\:\:\:\:\:\:\:\:\:\:\:\:\:\:\:\:\:\:\:\:\:\:
(d=2)\nonumber\\
&=&\frac{1}{\pi\sqrt{x^2-1}}\ln\bigg[\frac{x+\sqrt{x^2-1}}{x-\sqrt{x^2-1}}\bigg],
\:\:\:\:\:\:\: (d=1)
\label{F_parameter}
\end{eqnarray}
where $x=\frac{2p_F}{\kappa_{3,2}}$. If $\kappa_{3,2}\ll p_F$ then
$F\ll1$. For a detailed discussion the reader is referred to the
review article by Altshuler and Aronov.\cite{Altshuler}

\begin{figure}[t]
\includegraphics[scale=0.44]{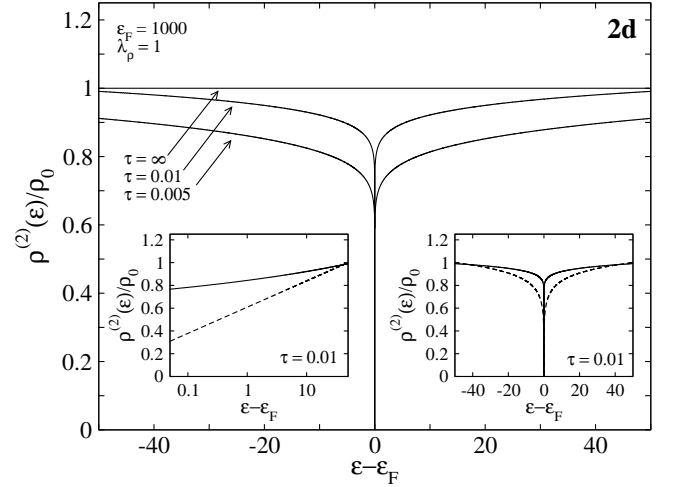}
\vspace*{-0.4cm} \caption{Normalized \emph{DOS} of a $2d$
disordered system as a function of the energy reckoned from the
Fermi level at $T=0$ for selected impurity concentrations. The
normalized pure \emph{DOS} $\rho_0^{(2)}$, corresponding to
$\tau=\infty$, is also given for comparison. In the inset we show
the comparison of the \emph{DOS} with the AAL theory (broken line)
in logarithmic scale (left hand side) and linear scale (right hand
side) for $\tau=0.01$. Notice the deviations from the logarithmic
behavior (broken line).} \label{fig3}
\end{figure}
\begin{figure}[!h]
\includegraphics[scale=0.44]{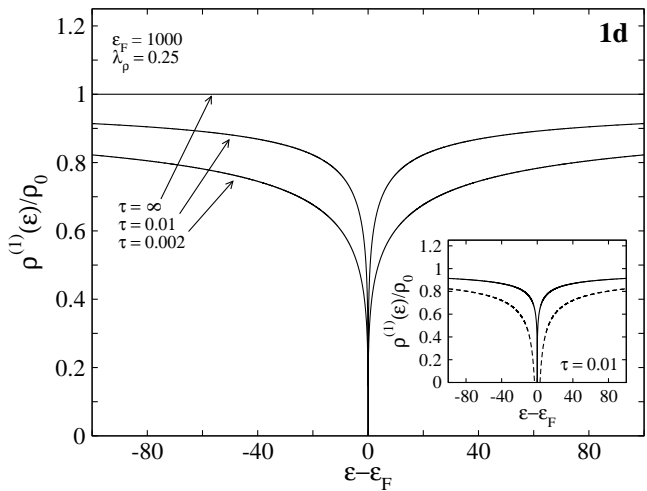}
\vspace*{-0.4cm} \caption{ Same as Fig.\ref{fig3} for quasi-$1d$
systems.} \label{fig4}
\end{figure}

So far we have restricted our attention here to the calculation of
self energies in the diffusion channel only. First order
corrections to the self energy given in Fig.~\ref{fig1} include
interactions in the Cooperon channel as well. The extension to
include the contribution of the Cooperon channel is
straightforward. It is clear that the additional self-energy parts
from this channel only renormalizes the dimensionless interaction
constant $\lambda_\rho$. Then, one can write the total self-energy
$\Sigma( {\bf p},i\epsilon_n)$ by adding the self energies in
diffusion $\Sigma^{D}({\bf p},i\epsilon_n)$ and Cooperon
$\Sigma^{C}({\bf p},i\epsilon_n)$ channels as $\Sigma ({\bf
p},i\epsilon_n)=\Sigma^{D}({\bf p},i\epsilon_n)+\Sigma^{C}({\bf
p},i\epsilon_n)$. Accordingly, renormalized dimensionless
interaction constant becomes $\lambda_\rho\Rightarrow
\lambda=\lambda^D_\rho+\lambda^C_\rho$, where $\lambda^C_\rho$ is
the interaction constant related to the self-energies in the
Cooperon channel in Fig.~\ref{fig1}. Consequently, replacing
$\lambda_\rho$ in Eqs.~(\ref{dos_2_d}) and (\ref{dos_1_d}) by
$\lambda$ we obtain the total contribution (including both
Cooperon and diffusion channels) to the \emph{DOS}.

In Figs.~\ref{fig3} and \ref{fig4} we present normalized
\emph{DOS} near the Fermi energy for a $2d$  and quasi-$1d$
systems at $T=0$ for selected impurity concentrations. In the
insets a comparison with the AAL theory is given. As seen, in
contrast to first-order perturbation theory, we obtain a vanishing
\emph{DOS} at Fermi level within the high-order perturbation
theory, \textit{i.e.}, in both dimensions
$\rho(\epsilon,T=0)\rightarrow 0$ for $\epsilon \rightarrow 0$.
This implies that interacting disordered $2d$ and quasi-$1d$
systems are in insulating state at $T=0$ in diffusive regime. On
the other hand, the suppression of the \emph{DOS} around Fermi
level is less singular compared to first-order perturbation
theory, for example in $2d$ systems the calculated \emph{DOS}
substantially deviates from the the logarithmic behavior (see
Fig.\ref{fig3}). Such behavior might be connected with the
excluded high-order self-energy diagrams in derivation of the
Eqs.~(\ref{dos_2_d}) and (\ref{dos_1_d}). Note that the complexity
of the diagrams increases dramatically  with increasing order of
the perturbation theory (For example see Fig.~\ref{fig2} for
second order self-energy contributions) and those neglected terms
can further decrease the \emph{DOS} around Fermi level. However,
these complex terms can not  be systematically included in our
treatment. It is worth to mention that within the present approach
the results of the AAL theory is recovered at small corrections,
\textit{i.e.}, when $\epsilon_F \tau \gg 1$ (see
Eqs.~(\ref{dos_2_d_aal}) and (\ref{dos_1_d_aal})). Note that in
low-dimensional systems the AAL theory is valid as long as
corrections to the \emph{DOS} are small. When corrections to the
\emph{DOS} are so large, one expects the perturbation calculation
to break down and the \emph{DOS} at the Fermi level diverges to
negative infinity.

Finally we should note that despite good agreement between present
results and non-perturbative studies for the energy and
temperature dependence of the \emph{DOS} at small corrections
(higher energies) an essential difference appears in the
asymptotic energy dependence of the \emph{DOS}. For example in
$2d$ systems (see Eq.~(\ref{dos_2_d})) our calculated zero
temperature \emph{DOS} vanishes as $\rho^{(2)}(\epsilon) \sim
[-\ln(\tau|\epsilon|)]^{-1/2}$ for $\epsilon \rightarrow 0 $,
whereas non-perturbative schemes give a power-law
behavior.\cite{Finkelstein,Kopietz,Kamenev}

\section{Zero-bias anomaly of the tunneling conductivity}
\label{section3}

The singularity in the energy dependence of the one-particle
\emph{DOS} would be reflected in thermodynamic and transport
properties of disordered conductors. A clear manifestation of this
effect is the minimum of the tunneling conductivity at zero bias.
The conductivity of the tunneling contact is related to \emph{DOS}
by
\begin{eqnarray}
\frac{\sigma^{(d)}(V,T)}{\sigma^{(d)}_0}&=&\frac{1}{4T}\int_{-\infty}^{+\infty}d\epsilon\frac{\rho^{(d)}
(\epsilon,T)}{\rho_0^{(d)}}\bigg[\frac{1}{\cosh^2\big(\frac{\epsilon-eV}{2T}\big)}
\nonumber \\
&&-\frac{1}{\cosh^2\big(\frac{\epsilon+eV}{2T}\big)}\bigg]
\label{tunneling_conductivity}
\end{eqnarray}
where $\sigma_0$ being classical conductivity, called Drude
expression. At zero temperature
Eq.\:(\ref{tunneling_conductivity}) reduces to
\begin{equation}
\frac{\sigma^{(d)}(V)}{\sigma^{(d)}_0}= \frac{\rho^{(d)}
(eV)}{\rho_0^{(d)}}
 \label{tunneling_conductivity2}
\end{equation}
As seen from Eq.\:(\ref{tunneling_conductivity2}) at $T=0$ the
$\sigma(V)$ is directly proportional to \emph{DOS}, thus the
measurement of the tunneling conductivity as a function of bias
voltage provides important information on the energy dependence of
the one-particle \emph{DOS}. On the other hand, at finite
temperatures the Eq.~(\ref{tunneling_conductivity}) can be written
as
\begin{eqnarray}
\frac{\sigma^{(d)}(T)}{\sigma^{(d)}_0}&=&\frac{2\rho^{(d)}(T)}{\rho_0^{(d)}}
\int_{0}^{1}\frac{dx}{\cosh^{2}x} \nonumber \\
&&+
\frac{2}{\rho_0^{(d)}}\int_{1}^{\infty}\frac{\rho^{(d)}(2Tx)}{\cosh^{2}x}
dx \label{tunneling_conductivity3}
\end{eqnarray}
In this expression  the major contribution comes from the first
integral and calculation for $2d$- and quasi-$1d$ systems gives
\begin{eqnarray}
\frac{\sigma^{(d)}(T)}{\sigma^{(d)}_0}&=&\frac{2C_{0}}{\sqrt{1+
\frac{2\lambda_\rho}{\epsilon_F\tau}\ln\big(\frac{1}{2\tau
T}\big)}}, \:\:\:\: (d=2)\nonumber \\
&=& \frac{2C_{0}}{\sqrt{1+\frac{2\lambda_\rho} {\sqrt{2\tau T}}}},
\:\:\:\:\:\:\:\:\:\:\:\:\:\:\:\:\:\: (d=1)
 \label{tunneling_conductivity4}
\end{eqnarray}
where $C_{0}\simeq 1$ is a coefficient. The expression
(\ref{tunneling_conductivity4}) shows the change of the tunneling
conductivity of the  low-dimensional systems  corresponding the
the temperature in the small value of the potential ($V
\rightarrow 0 $).

Since early 1980's several research groups have performed
tunneling measurements of quasi-$2d$ disordered metal and
semi-metal films and observed the predicted logarithmic dependence
of tunneling conductance with
voltage.\cite{tunnel_2,tunnel_3,tunnel_4,tunnel_5,tunnel_6} Note,
however that the experimental techniques used in these early
studies was not capable of detecting tunneling current for small
voltages. Access to such regimes  becomes possible only recently
with TDCS method as we mentioned in the introduction. It should be
emphasized that TDCS is unique in allowing complete extraction of
the tunneling spectrum of low-dimensional systems. Using this
technique Chan \textit{et al}., for the first time measured the
entire voltage dependence of the tunneling conductance of a $2d$
electron system in a GaAs/AlGaAs heterostructure for various
electron densities.\cite{tdcs_2,tdcs_3} The authors observed the
expected logarithmic Coulomb anomaly only in the case of a small
suppression of the tunneling current ($\epsilon_F \tau \gg 1$).
However, for large suppressions corresponding to small electron
densities for which $\epsilon_F \tau \sim 1$ the functional form
of the ZBA vs bias voltage was significantly deviating from the
predictions of the AAL theory, especially  in the regime of very
small voltages. Furthermore, application of magnetic field
perpendicular to the $2d$ plane results in a linear dependence of
the tunneling conductance on voltage near zero bias for all
magnetic field strengths and electron densities. This latter
phenomena is not yet completely understood.\cite{tdcs_2} Peculiar
behavior of the tunneling conductivity at small electron densities
can be qualitative explained by the present theory. Indeed, as
seen from Fig.~\ref{fig3} the \emph{DOS} of $2d$ systems strongly
deviates from the logarithmic behavior for small energies in
agreement with observations of Chan \textit{et al}. However, in
our case the deviations seem to be stronger than the one observed
in the experiment. As commented in previous section this might be
due to excluded high-order self-energy diagrams in calculation of
the corrections to the \emph{DOS}.

In contrast to $2d$ systems,  ZBA in quasi-$1d$ conductors
received less attention. White \textit{et al.}, reported  the
first systematic study of the corrections to the \emph{DOS} of
quasi-$1d$ granular aluminum wires.\cite{White} The obtained
corrections to the \emph{DOS} somehow  do not have the $V^{-1/2}$
dependence predicted by AAL theory, but are significantly larger
than the corrections observed in corresponding bulk samples.
Pierre \emph{et al.}, measured the tunneling \emph{DOS} of a
metallic wire in perturbative regime in a controlled way and
obtained the predicted behavior for the suppression of the
tunneling conductance.\cite{Pierre} Recently, Yu and Natelson
studied the ZBA in electrochemically fabricated disordered
nanojunctions of various size. For large junctions the authors
obtained  a small ZBA which is consistent with the perturbative
theory of AAL. \cite{Natelson_1} However, in atomic scale
junctions the observed ZBA was approaching 100\% conductance
suppression as $T \rightarrow 0$.\cite{Natelson_2}

Finally, we will briefly discuss the ZBA observed in doped
multiwall carbon nanotubes (MWCNT). MWCNT constitute a quasi-$1d$
systems with fascinating physical properties. Several experiments
have demonstrated that the charge transport in these systems is
diffusive\cite{Langer,Bachtold_1,Bachtold_2,Tarkiainen} i.e.,
showing typical weak localization features in the
magnetoconductance and thus, their physical properties show
Fermi-liquid (FL) behavior. However, the functional form of the
observed ZBA in MWCNT is characteristics of the Luttinger-liquid
state in $1d$ clean (ballistic) systems of interacting electrons.
One of the main features of the LL state in $1d$ is the power-law
dependence of physical quantities, for instance tunneling
\emph{DOS}, as a function of energy or temperature ($\rho
(\epsilon) \sim \epsilon^{\alpha}$).\cite{LL_Kane} In  MWCNT the
observed values for the exponent $\alpha$ is rather scattered
between 0.04 and 0.37 depending on the geometry of the
samples.\cite{alpha_1,alpha_2,alpha_3,alpha_4,alpha_5,alpha_6,alpha_7}
The observed peculiar behavior of the tunneling conductivity was
attributed to the disorder enhanced $ee$ interaction effects and
its theoretical descriptions was beyond the first-order
perturbation theory due to large suppressions of the tunneling
conductance. Thus, a non-perturbative treatment has recently been
put forward by Egger and Gogolin.\cite{Egger} The authors
predicted a geometry dependent LL-like ZBA in doped MWCNT. Somehow
the situation is not so different within present scheme, it
follows from the Eq.~(\ref{dos_1_d}) that the tunneling \emph{DOS}
around Fermi level for quasi-$1d$ systems presents a power-law
behavior ($\rho^{(1)}(\epsilon) \sim (\epsilon\tau)^{1/4}$) with
an exponent $\alpha=0.25$. Note that in several tunneling
experiments on doped MWCNT the observed value of $\alpha$ is close
to $0.25$ in good agreement with our
predictions.\cite{alpha_1,alpha_6,alpha_7}

\section{Conclusions}
\label{section4}

In conclusion, we propose a diagrammatic approach to study
critical behavior of the  one-particle \emph{DOS} of
low-dimensional disordered  metals in diffusive regime. By a
geometric re-summation of the most singular first order
self-energy corrections via the Dyson equation we obtain a
non-divergent solution for the \emph{DOS} at low energies, while
for higher energies the well-known Altshuler-Aronov corrections
are recovered. At the Fermi level
$\rho^{(d)}(\epsilon,T=0)\rightarrow 0$, this indicates that
interacting disordered $2d$ and quasi-$1d$ systems are in
insulating state at zero temperature. However, asymptotic energy
dependence of the calculated \emph{DOS} differs from those
obtained by non-perturbative methods. For $2d$ systems at zero
temperature the \emph{DOS} vanishes as $\rho^{(2)}(\epsilon) \sim
[-\ln(|\epsilon|\tau)]^{-1/2}$ for $\epsilon \rightarrow 0 $,
whereas non-perturbative schemes give a power-law behavior (See
Refs.~\onlinecite{Finkelstein,Kopietz} and \onlinecite{Kamenev}).
In contrast  to $2d$ case, a power-law behavior
($\rho^{(1)}(\epsilon) \sim (\epsilon\tau)^{1/4}$) is predicted
for the asymptotic energy dependence of the \emph{DOS} of
quasi-$1d$ systems. The obtained results are in good agreement
with recent tunneling experiments on two-dimensional GaAs/AlGaAs
heterostructures and quasi-one-dimensional doped multiwall carbon
nanotubes.

\end{document}